\documentclass[twocolumn,amsmath,amssymb,pre]{revtex4-1}

\usepackage{graphicx}
\usepackage{color}
\usepackage{dcolumn}
\usepackage{amsmath}
\usepackage{subfigure}
\usepackage{sidecap}

\usepackage{mathrsfs}
\usepackage{amsfonts}
\usepackage{indentfirst}
\usepackage{bbm}
\usepackage{dsfont}
\usepackage{makecell}

\usepackage[colorlinks,citecolor=blue]{hyperref}

\newcommand{\be}{\begin{equation}}
\newcommand{\ee}{\end{equation}}
\newcommand{\bey}{\begin{eqnarray}}
\newcommand{\eey}{\end{eqnarray}}
\newcommand{\bw}{\begin{widetext}}
\newcommand{\ew}{\end{widetext}}

\newcommand{\ra}{\rangle}
\newcommand{\la}{\langle}

\newcommand{\ba}{\begin{array}}
\newcommand{\ea}{\end{array}}
\newcommand{\bi}{\begin{itemize}}
\newcommand{\ei}{\end{itemize}}
\newcommand{\bem}{\begin{enumerate}}
\newcommand{\eem}{\end{enumerate}}

\begin{document}

\title{Excited stated quantum phase transitions and the entropy of the work distribution in the
anharmonic Lipkin-Meshkov-Glick model}

\author{Haiting Zhang$^{1}$}
\author{Yifan Qian$^{1}$}
\author{Zhen-Xia Niu$^{1}$}\email{niuzhx@zjnu.edu.cn}
\author{Qian Wang$^{1,2}$}\email{qwang@zjnu.edu.cn}

\affiliation{$^{1}$Department of Physics, Zhejiang Normal University, Jinhua 321004, China \\
$^2$CAMTP-Center for Applied Mathematics and Theoretical Physics, University of Maribor,
Mladinska 3, SI-2000, Maribor, Slovenia}

\date{\today}

\begin{abstract}

Studying the implications and characterizations of the 
excited state quantum phase transitions (ESQPTs) would enable us to understand 
various phenomena observed in quantum many body systems.
In this work, we delve into the affects and characterizations of the ESQPTs 
in the anharmonic Lipkin-Meshkov-Glick (LMG) model 
by means of the entropy of the quantum work distribution. 
The entropy of the work distribution measures the complexity of the work distribution 
and behaves as a valuable tool for analyzing nonequilibrium work statistics.
We show that the entropy of the work distribution captures salient signatures of the underlying 
ESQPTs in the model.
In particular, a detailed analyses of the scaling behavior of the maximal entropy verifies that
it acts as a witness of the ESQPTs. 
We further demonstrate that the entropy of the work distribution also reveals the 
features of the ESQPTs in the energy space and can be used to determine their critical energies. 
 Our results provide further evidence of the usefulness of the entropy of the work distribution
 for investigating various phase transitions in quantum many body systems and
 open up a promising way for experimentally exploring the signatures of ESQPTs.

\end{abstract}

\maketitle

\section{Introduction}

Excited state quantum phase transitions (ESQPTs) 
\cite{Cejnar2006,Cejnar2008,Caprio2008,Stransky2014,Cejnar2021} are a generalization 
of the ground state quantum phase transitions (QPTs) \cite{Carr2010,Sachdev2011} and 
have triggered a numerous of investigations
in understanding their effects and signatures in
a wide variety of quantum many body systems,
such as the Dicke model \cite{Brandes2013,Magnani2014,Lobez2016,Kloc2018,Corps2021}, 
the Rabi model \cite{Puebla2016,Stransky2021}, 
the periodically driving systems \cite{Bastidas2014,Saiz2023}, 
the spinor Bose-Einstein condensates \cite{Feldmann2021,Cabedo2021,NiuZ2023}, 
the Lipkin-Meshkov-Glick (LMG) model 
\cite{Caprio2008,YuanZG2012,Engelhardt2015,Kopylov2015,Santos2015,Sindelka2017,
WangQ2019,Corps2022,Santos2016,WangQ2021},
and Kerr nonlinear oscillator \cite{Chavez2022,WangQW2020}, to name a few.
In particular, ESQPTs have been experimentally observed in the
superconducting microwave billiards \cite{Dietz2013} and the quantum gas \cite{Meyer2023}.
ESQPTs are usually characterized by the singularities in the density of states
\cite{Caprio2008,Stransky2014,Cejnar2021} and
play an important role in a diverse range of situations, including 
decoherence process \cite{Relano2008,PerezF2009},
quantum quench dynamics \cite{Kloc2018,Santos2016,
Chavez2022,NiuZ2023,PerezF2017,WangQa2019,Pilatowsky2020,Kloc2021,WangQ2021}, 
quantum chaos \cite{Lobez2016,PerezF2011,Garca2021}, 
quantum metrology \cite{ZhouL2023}, and dynamical tunneling \cite{Nader2021}
as well as isomerization reactions \cite{Khalouf2019}.
Moreover, the efforts to identify the order parameters for ESQPTs \cite{Corps2021,WangQa2019} 
provide further understanding on their properties.
Different aspects of ESQPTs have been reviewed in detail in Ref.~\cite{Cejnar2021}.

Recently, the anharmonicity induced ESQPTs have attracted a lot of attention
\cite{Khalouf2019,PerezB2010,Khalouf2022b,Gamito2022,Khalouf2022a}.
In contrast to the usual ESQPTs \cite{Caprio2008,Stransky2014,Cejnar2021}, 
which are associated with the ground state QPTs,
the anharmonicity triggered ESQPTs are independent of the ground state QPTs
and have different physical origin as compared to the usual ones.
It has been demonstrated that the onset of anhamonicity induced ESQPTs is a consequence of
the changes of the boundary in the finite dimensional Hilbert space 
of the system \cite{Khalouf2022b,Gamito2022}.    
It is worth mentioning that the finitness of the Hilbert space of the system also results in 
another kind of ESQPTs, known as the static ESQPTs \cite{Magnani2014}, 
which have no impacts on the system dynamics.
The differences between the static and anharmonicity induced ESQPTs 
have been pointed out in Ref.~\cite{Gamito2022}.
Both static and dynamical aspects of the anharmonicity induced ESQPTs have been throughly
investigated in several systems \cite{Khalouf2022b,Gamito2022,Khalouf2022a}.
However, more works are still required for a better understanding of their 
affects and associated signatures.

The aim of the present work is to extend the studies on the effects of the anharmonicity induced
ESQPTs on the system static and dynamical properties to the nonequilibrium thermodynamics. 
To this end, we carry out a detailed investigation on the work statistics of the 
anharmonic LMG model \cite{Khalouf2022a,Gamito2022} in a sudden
quench process by means of the entropy of the quantum work distribution.
As a measure of the complexity of the quantum work distribution, the work distribution entropy
conveys a richness of information of the statistics of nonequilibrium quantum work, 
and particularly it can be used as a powerful tool to diagnose the localization transition 
in many body systems \cite{Kiely2023}.

The anharmonic LMG model is obtained by including an anharmonic term 
in the LMG model \cite{Gamito2022}.
It was shown that in addition to the ESQPT that has been observed in the LMG model, the inclusion
of the anharmonic term triggers a new ESQPT \cite{Gamito2022}. 
We demonstrate that both ESQPTs are characterized by the logarithmic divergence 
of the density of states at different critical energies.
By analyzing the classical limit of the model, we perform a detailed analysis of the two 
ESQPTs and show how to understand them from the changes of the avaliable phase space volume.  
To take the system out of the equilibrium state, we employ the sudden quench process.
We show that the underlying ESQPTs leave a strong imprint in 
the nonequilibrium work statistics, 
resulting in a notable change in the behavior of the work distribution entropy.
The peak exhibited by the work distribution entropy 
around the critical points of ESQPTs leads us to identify it as the proxy for both ESQPTs.
We confirm this by performing a detailed scaling analyses on the maximal entropy.
Furthermore, for the sake of completeness, we also discuss the energy dependence of the work
distribution entropy and illustrate that it behaves as a useful tool to 
detect the critical energies of ESQPTs.

The remainder of the article is structured as follows.
In Sec.~\ref{SecondS}, we beirfly review the definition and several features of 
the entropy of the work distribution.
In Sec.~\ref{ThirdS}, we describe the anharmonic LMG model, 
discuss the signatures of ESQPTs in the model and study its classical counterpart. 
We analyze the physical origins of ESQPTs and obtain explicit expressions of their critical energies.
In the following Sec.~\ref{FourthS}, we report our main results, show how 
the ESQPTs get reflected in the behavior of the work distribution entropy.
We finally summarize our findings and discuss their potential extensions in Sec.~\ref{FivthS}.

\section{Entropy of the work distribution}\label{SecondS}

Let us consider an isolated quantum system subjected to an external time-dependent driving 
of the control parameter in a time interval $[0, \tau]$. 
The system is initinally prepared in a generic state $\rho_i$ with 
inital Hamiltonian $H_i=\sum_nE_n^i|n_i\ra\la n_i|$. 
Here, $|n_i\ra$ is the $n$th eigenstate of $H_i$ and $E_n^i$ is its corresponded eigenvalue.
After the driving, the state of the system becomes $\rho_f={U}\rho_i{U}^\dag$ where $U$ represents 
the unitary evolution. 
The driving process also changes system Hamiltonian from $H_i$ to the final one
$H_f=\sum_kE_k^f|k_f\ra\la k_f|$, where $E_k^f$ is the 
eigenvalue of the $k$th eigenstate, $|k_f\ra$, of $H_f$.
The work done during this process is a random variable and
is given by the energy difference between the final and initial Hamiltonians. 
Thus, in order to understand the nonequilibrium thermodynamics in the isolated quantum system,
one needs to consider the work distribution rather than the work itself.

There are several different forms of the work distribution
\cite{Solinas2015,Sampaio2018,Francica2022a,
Francica2022b,Lostaglio2022,Cerisola2023,Talkner2007}, 
depending on which scheme that is employed to evaluate the work during the process.
Among them, we focus on the work distribution defined following the two point measurement
scheme \cite{Talkner2007,Tasaki2000,Kurchan2001}, which is a most popular scheme for studying
of the work statistics in driven isolated systems \cite{Deffner2008,Silva2008,Doner2012,
Marino2014,Campbell2016,Goold2018,FeiZ2020,Zawadzki2023}.
The two point measurement scheme consists of measuring the energy of the system at the begining
and at the end of the driving.  
Accordingly, the corresponding works, $W=E_k^f-E_n^i$, are distributed as
\be \label{DefPw}
  P(W)=\sum_{n,k}p_{k,n}\delta[W-(E_k^f-E_n^i)],
\ee
where $p_{k,n}$ denots the joint probability of the two energy measurements and is given by 
\be \label{JointPrb}
  p_{k,n}=\mathrm{Tr}[U\Pi_i^{(n)}\rho_i\Pi_i^{(n)}U^\dag\Pi_f^{(k)}],
\ee
with $\Pi_i^{(n)}=|n_i\ra\la n_i|$ and  $\Pi_f^{(k)}=|k_f\ra\la k_f|$.

The work distribution for a quantum many-body system is complex.
Hence, one usually studies the moments or cumulants of work, such as the mean and variance, 
in analyzing the work statistics of quantum systems
\cite{Silva2008,Doner2012,Marino2014,Zawadzki2023,Zawadzki2020,Mzaouali2021,
Fusco2014,QianW2017}. 
Although the work moments or cumulants can capture several features of 
nonequilibrium thermodynamics in a variety of many-body systems, 
they cannot reveal the full information included in the work distribution.
Very recently, a quantity that measures the complexity of $P(W)$ has been introduced, that is,
the entropy of $P(W)$ \cite{Kiely2023}:
\be\label{EntropyPw}
  S_W=-\sum_WP(W)\ln[P(W)].
\ee
It is easy to see that $S_W\in[0, \ln\mathcal{D}^2]$ with $\mathcal{D}$ being 
the Hilbert space dimension of the quantum system.
If the work is deterministic, then $S_W=0$, 
while $S_W=\ln\mathcal{D}^2$ implies $P(W)$ is uniform.

It has been demonstrated that the entropy $S_W$ unveils a richness information of the work distribution
and provides a useful tool for understanding the nonequilibrium 
thermodynamics in quantum systems \cite{Kiely2023}. 
Moreover, the entropy also acts as a sensitive probe of the localization transition.
In this work, we use the entropy $S_W$ to explore the implications of the ESQPT 
on the nonequilibrium thermodynamics
in the anharmonic LMG model.
We also examine whether the usefuless of the entropy in detecting 
the presence of the localization transition can be extended to the case of ESQPT.

  \begin{figure}
  \includegraphics[width=\columnwidth]{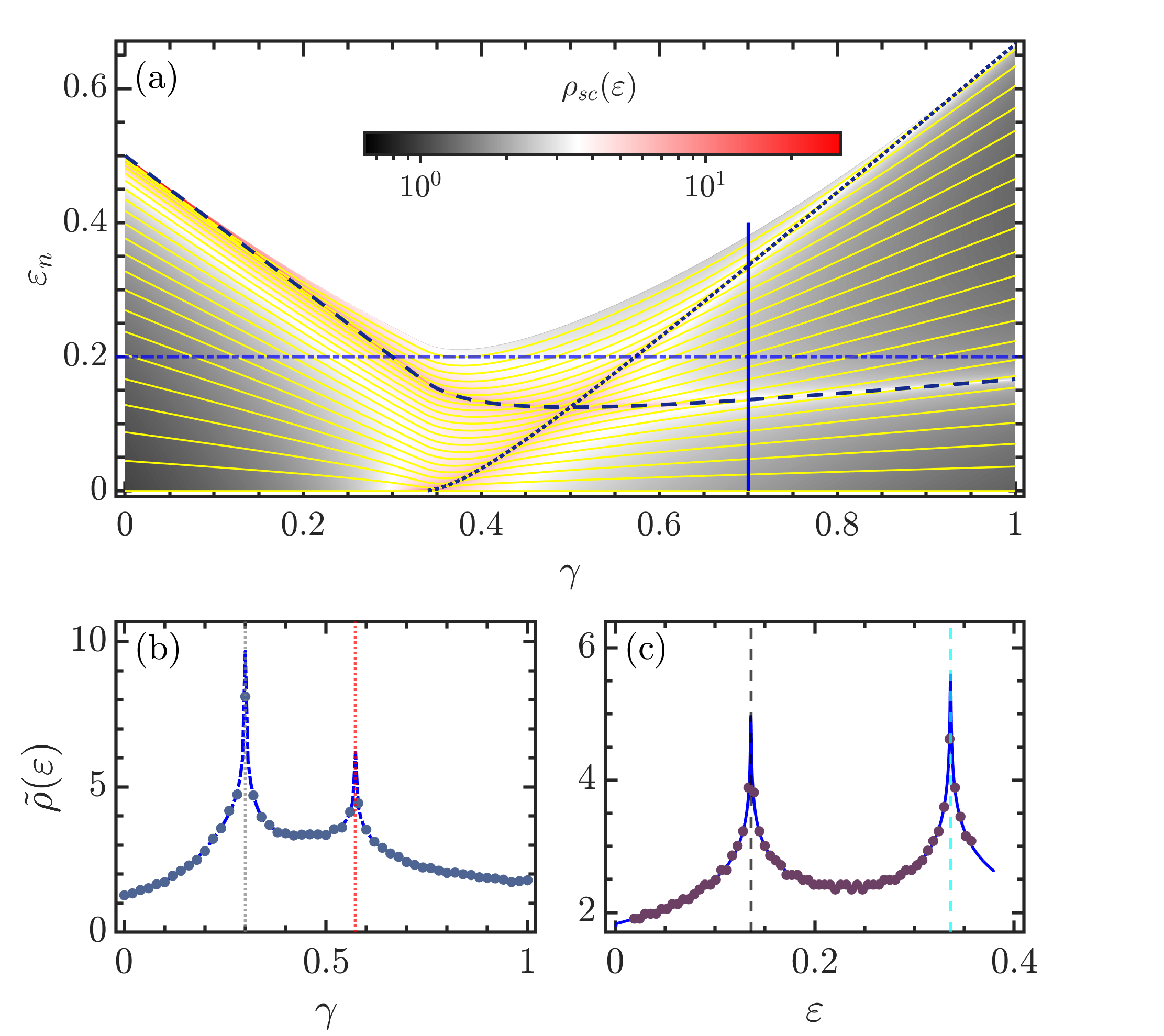}
  \caption{(a) Rescaled even parity energy levels $\varepsilon_n=(E_n-E_0)/N$ 
  (yellow solid lines and increasing in step of $\Delta n=3$) 
  of the anharmonic LMG model as a function of the control 
  parameter $\gamma$ for a system with $N=2j=130$.
  Heat map plot depicting the semiclassical approximation of the
  density of states $\rho_{sc}(\varepsilon)$ [cf.~Eq.(\ref{DoS})] as a function 
  of $\varepsilon_n$ and $\gamma$.
  The dark blue dotted and dashed curves indicate the ESQPTs critical energy $\varepsilon_{c,1}$
  and $\varepsilon_{c,2}$, respectively. 
  The horizontal blue dot dashed line marks $\varepsilon=0.2$, while the vertical blue solid line
  indicates $\gamma=0.7$.
  (b) Rescaled quantum density of states $\tilde{\rho}(\varepsilon)=\rho(E)/N$ as a function of
  $\gamma$ along $\varepsilon=0.2$ for the system size $N=2j=5000$.
  The blue dot dashed curve denotes $\rho_{sc}(\varepsilon)$, obtained from Eq.~(\ref{DoS}).
  The vertical red and gray dotted lines are the critical $\gamma$ values of the ESQPTs 
  provided by the equations $\varepsilon_{c,1}=\varepsilon_{c,2}=0.2$. 
 (c) Dependence of $\tilde{\rho}(\varepsilon)$ on $\varepsilon$ along $\gamma=0.7$
 for the same system size as in panel (b).
 The blue solid curve represents $\rho_{sc}(\varepsilon)$ in Eq.~(\ref{DoS}), while the vertical
 cyan and black dashed lines are respective marks $\varepsilon_{c,1}$ and $\varepsilon_{c,2}$. 
  In all panels, the anharmornicity parameter $\alpha=0.5$.  
  All quantities are unitless.}
  \label{EgDoS}
 \end{figure}

\section{Model}\label{ThirdS}

As a generalization of the well known LMG model \cite{Lipkin1965a,Meshkov1965b,Glick1965c}, 
which describes $N$ spin-$1/2$ mutual interacting
particles in an external field and was widely studied in various areas
\cite{Dusuel2004,Dusuel2005,Castanos2006,Ribeiro2007,Ribeiro2008,
Campbell2015,Russomanno2017,ZhangZ2022,
Afrasiar2022,Kumari2022,Zibold2010,Muniz2020,KaiX2020}, 
the anharmonic LMG model includes an anharmonic term and its Hamiltonian reads ($\hbar=1$) \cite{Khalouf2022a,Gamito2022}
\begin{align}\label{ALMGH}
   H=&\frac{2\gamma}{N}(J^2-J_x^2)+(1-\gamma)\left(J_z+\frac{N}{2}\right) \notag \\
     &-\frac{\alpha}{N}\left(J_z+\frac{N}{2}\right)\left(J_z+\frac{N}{2}+1\right),
\end{align}
where $J_{x,y,z}=\sum_{i=1}^N\sigma_i^{x,y,z}$ are the collective spin operators with
$\sigma_i^{x,y,z}$ denote the $i$th spin Pauli matricies, $\gamma\in[0,1]$ is the control parameter,
and $\alpha>0$ represents the strength of the anharmonic effect.
The Hamiltonian (\ref{ALMGH}) reduces to the original LMG model when $\alpha=0$,
while it includes the interactions between spins along $z$ direction for $\alpha\neq0$ cases.

 \begin{figure*}
  \includegraphics[width=\textwidth]{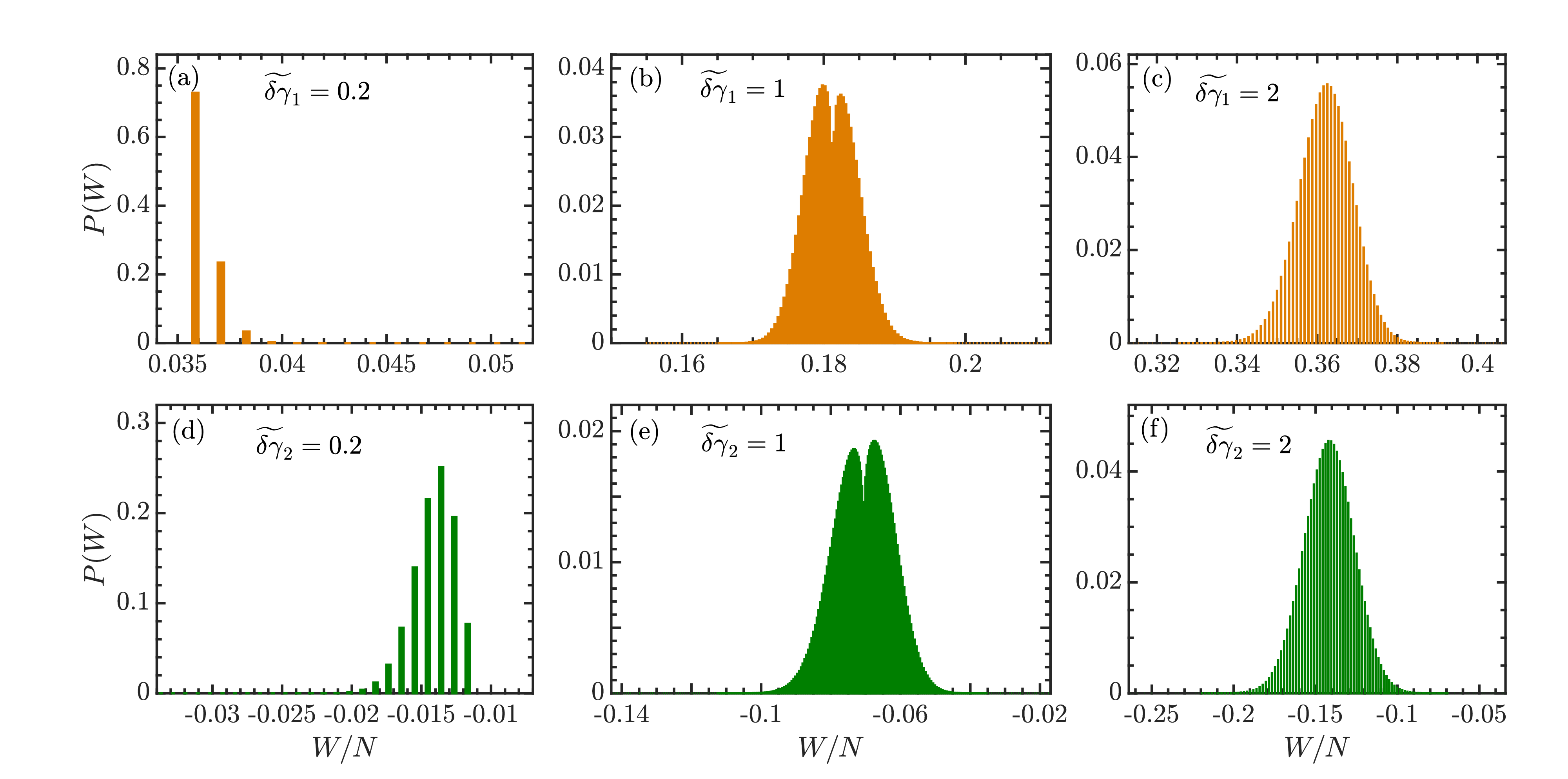}
  \caption{Work distributions of the anharmonic LMG model (\ref{ALMGH}).
  (a)-(c) $P(W)$ for the first ESQPT with different values of the rescaled quenching strength
  $\widetilde{\delta\gamma}_1=\delta\gamma/\delta\gamma_{c,1}$ and initial state is given by the
  ground state of $H_i$. 
  Here, $\delta\gamma_{c,1}$ is obtained from Eq.~(\ref{CrtFirst}).
  (d)-(f) $P(W)$ for the second ESQPT with several values of 
  $\widetilde{\delta\gamma}_2=\delta\gamma/\delta\gamma_{c,2}$ and the initial state 
  is the highest excited eigenstate of $H_i$.
  Here, $\delta\gamma_{c,1}$ is given by Eq.~(\ref{CrtSecond}).
  In all panels, $\gamma_i=0.7$, the anharmornicity parameter $\alpha=0.5$,
  and the system size $N=2j=800$.  
  All quantities are unitless.}
  \label{PdfWork}
 \end{figure*}

One can easily show that the total spin $J^2=J_x^2+J_y^2+J_z^2$ commutes with the 
Hamiltonian (\ref{ALMGH}).
Hence, we focus on the $j=N/2$ sector with the Hilbert space dimension 
$\mathcal{D}_\mathscr{H}=N+1$. 
Moreover, the conservation of the parity $\Pi=e^{i\pi(j+J_z)}$ further allows us
to split the Hilbert space into two subspaces, one has even parity with dimension 
$\mathcal{D}_\mathscr{H}^e=N/2+1$ and the other is the odd parity subspace with dimension 
$\mathcal{D}_\mathscr{H}^o=N/2$.
In this work, we restrict to the even parity subspace, which includes the ground state of the model.

It is known that the anharmonic LMG model undergoes 
a second-order ground state QPT at $\gamma_c=1/3$
and two ESQPTs, revealed by the clustering of the eigenlevels \cite{Gamito2022,Khalouf2022a}.
In Fig.~\ref{EgDoS}(a), we plot how the rescaled excitation energies 
$\varepsilon_n=(E_n-E_0)/N$ (white solid lines) vary as a function of 
the control parameter $\gamma$ for a system with $N=130$ and $\alpha=0.5$.
Here, $E_n$ is the $n$th energy level with $n=0$ being the ground state.
Clearly, the energy spectrum of anharmonic LMG is more complex.
One can see that the eigenlevels cluster along two different lines, which indicate 
the exsitence of two ESQPTs.
The first ESQPT, marked by the dark blue dotted line, is the same as the one in the LMG model
\cite{Santos2015,Santos2016,PerezF2017} and only appears for $\gamma>\gamma_c=1/3$,
while the second one, triggered by the anharmonic term, 
can present in the entire range $\gamma\in[0,1]$, as marked by the dark blue 
dashed line in Fig.~\ref{EgDoS}(a).
The clustering of the eigenlevels implies that
both ESQPTs are characterized by a high density of states, $\rho(E)=\sum_n\delta(E-E_n)$. 
This is visible in Figs.~\ref{EgDoS}(b) and \ref{EgDoS}(c),
where we plot the rescaled density of states, $\tilde{\rho}(\varepsilon)=\rho(E)/N$, as a function of 
$\gamma$ and $\varepsilon$, respectively, for a system size $N=5000$ and $\alpha=0.5$. 
It is easy to see that near the critical points of ESQPTs
the density of states exhibits two remarkable peaks
which will translate into the logarithmic divergences in the thermodynamic limit \cite{Santos2016,Gamito2022}.
We would like to point out that the second ESQPT has strong impacts 
on system dynamics \cite{Khalouf2022a},
in contrast to the static ESQPT introduced in Ref.~\cite{Magnani2014}. 

The onset of ESQPT is closely connected to the changes of the available phase space in the underlying
classical systems \cite{Caprio2008,Stransky2014,Cejnar2021}. 
We therefore explore the classical limit of the anharmonic LMG model (\ref{ALMGH}), in
order to get a better understand the signatures of both ESQPTs.

\subsection*{Classical limit of the model}
 
The classical counterpart of the Hamiltonian (\ref{ALMGH}) can be obtained by employing the
generalized $\mathrm{SU}(2)$ spin coherent states, which are defined as
\cite{Radcliffe1971,ZhangW1990,Gazeau2009,Perelomov2012}  
\begin{align}
  |\xi\ra&=\frac{\exp(\xi J_-)}{(1+|\xi|^2)^j}|j,j\ra, \notag \\
    &=\sum_{m=-j}^{m=j}\frac{\xi^{j-m}}{(1+|\xi|^2)^j}\sqrt{\frac{(2j)!}{(j+m)!(j-m)!}}|j,m\ra.
\end{align}
Here, the complex parameter $\xi=(q+ip)/\sqrt{4-p^2-q^2}$ 
with $\{(p,q)|p^2+q^2\leq4\}$ are the canonical variables, 
$J_\pm=J_x\pm iJ_y$ denote the spin ladder operators, and $J_z|j,j\ra=j|j,j\ra$.
The classical counterpart of the Hamiltonian (\ref{ALMGH}) is obtained from
its expectation value with respect to the coherent state 
and normalizing by the system size in the classical limit ($N\to\infty$).
Employing the relations \cite{Radcliffe1971}
\begin{align}
  \la\xi|J_z|\xi\ra&=j\left(\frac{|\xi|^2-1}{|\xi|^2+1}\right),\notag \\
  \la\xi|J_-|\xi\ra&=\la\xi|J_+|\xi\ra^\ast=\frac{2j\xi}{|\xi|^2+1},
\end{align}
after some algebra, it is straightforward to find that the classical Hamiltonian can be written as
\begin{align}\label{ClassicalH}
   \mathcal{H}_c(p,q)
    =&\frac{1-\gamma}{4}(p^2+q^2)-\frac{\gamma}{8}q^2(4-p^2-q^2) \notag \\
        &-\frac{\alpha}{16}(p^2+q^2)^2+\frac{\gamma}{2}.
\end{align} 
The associated classical equations of motion are
\begin{align}\label{EMO}
  \dot{q}=\frac{\partial\mathcal{H}_c(p,q)}{\partial p} 
     =&\frac{1-\gamma}{2}p+\frac{\gamma}{4}q^2p-\frac{\alpha}{4}p(p^2+q^2),\notag \\
  \dot{p}=-\frac{\partial\mathcal{H}_c(p,q)}{\partial q}
     =&-\frac{1-\gamma}{2}q+\frac{\gamma}{4}q(4-p^2-2q^2) \notag \\
     &+\frac{\alpha}{4}q(p^2+q^2).
\end{align}

The phase space structure of a classical model is determined by its fixed points 
\cite{Stransky2014,Cejnar2021}, which are coincided with the stationary points of the dynamics. 
Nullification of equations of motion (\ref{EMO}) results in different fixed points depending on the
$\gamma$ and $\alpha$ values. 
For $\gamma<\gamma_c=1/3$, the classical system has a fixed point $(p_0,q_0)=(0,0)$, 
corresponding to the energy $\mathcal{E}_0=\gamma/2$. 
This is the minimal energy of the system when $\gamma<\gamma_c$
However, for $\gamma\geq\gamma_c=1/3$, two additional fixed points 
$(p_1,q_1)=(0,\pm\sqrt{2(3\gamma-1)/(2\gamma-\alpha)})$ appear with the energy
$\mathcal{E}_1= \gamma/2-(3\gamma-1)^2/(8\gamma-4\alpha)$ 
and the previous fixed point becomes the saddle point,
which leads to the separatrix in classical dynamics and defines an ESQPT.
The energy difference $\mathcal{E}_1-\mathcal{E}_0$ defines the equation of the separatix and 
provides the ESQPT critical energy,
\be\label{CrtEg1}
   \varepsilon_{c,1}=\mathcal{E}_0-\mathcal{E}_1=\frac{(3\gamma-1)^2}{4(2\gamma-\alpha)}, \quad
     \gamma\geq\gamma_c,
\ee  
which is plotted as the dark blue dotted line in Fig.~\ref{EgDoS}(a).

Apart from above mentioned fixed points, the anharmonicity gives rise to other fixed points
$(p_2,q_2)=(\pm\sqrt{(2\gamma+2-4\alpha)/\gamma},\pm\sqrt{(4\alpha-2+2\gamma)/\gamma})$
with associated energy $\mathcal{E}_2=1-\gamma/2-\alpha$, as long as $\gamma\geq|1-2\alpha|$. 
The existence of these fixed points reflects that there is a second separatrix in the classcial dynamics.
This means the onset of a new ESQPT with the critical excitation energy,
\begin{align}\label{CrtEg2}
  \varepsilon_{c,2}=
  \left\{
  \begin{aligned}
  &\mathcal{E}_2-\mathcal{E}_0=1-\gamma-\alpha,\ \gamma<\gamma_c,  \\
  &\mathcal{E}_2-\mathcal{E}_1=\frac{(1+\gamma-2\alpha)^2}{4(2\gamma-\alpha)},\ 
  \gamma\geq\gamma_c,
  \end{aligned}
  \right.
\end{align}
which is marked with the dark blue dashed line in Fig.~\ref{EgDoS}(a). 
As can be easily seen in Eq.~(\ref{CrtEg2}), the critical energy $\varepsilon_{c,2}$ 
exists in the full range of $\gamma$ value, unlike $\varepsilon_{c,1}$, which
can only be found in the broken-symmetry phase $\gamma\in(\gamma_c,1]$. 

 \begin{figure}
  \includegraphics[width=\columnwidth]{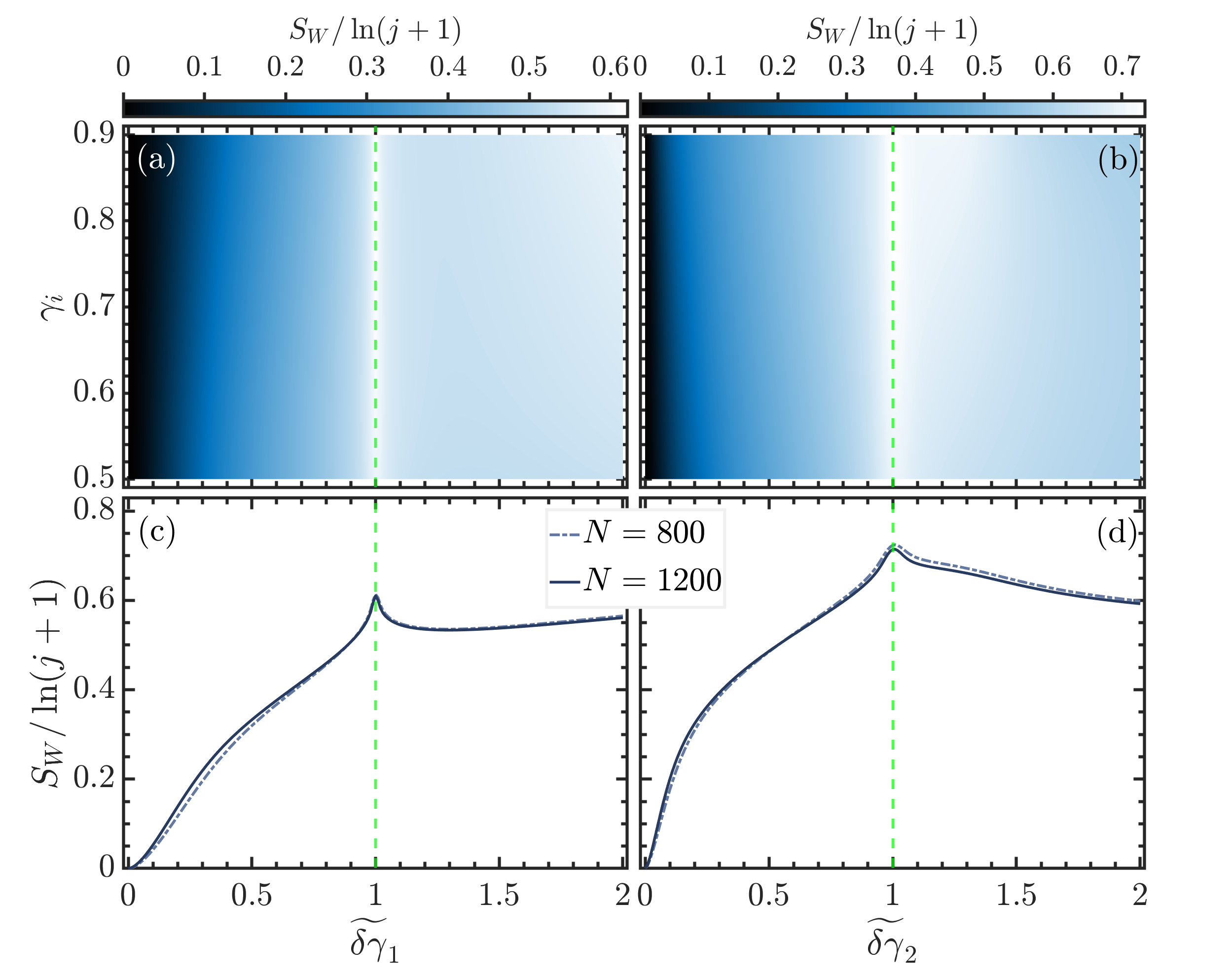}
  \caption{(a)-(b) Rescaled entropy of $P(W)$ as a function of 
  $\widetilde{\delta\gamma}_a=\delta\gamma/\delta\gamma_{c,a} (a=1,2)$  and $\gamma_i$
  for (a) the first ESQPT and (b) the second ESQPT with $N=2j=800$. 
  Here, $\delta\gamma_{c,a}$ are obtained from Eqs.~(\ref{CrtFirst}) and (\ref{CrtSecond}).
  (c)-(d) Rescaled entropy of $P(W)$ as a function of $\widetilde{\delta\gamma}_a$ 
  for several system sizes, for (c) the first ESQPT and (d) the second ESQPT with $\gamma_i=0.7$.
  The vertical green dashed line in each panel marks the critical quenching strengths 
  $\delta\gamma_{c,a}$.
  In all panels, the anharmornicity parameter $\alpha=0.5$
  All quantities are unitless.}
  \label{PWentropy}
 \end{figure}

The presence of the separatrices in the classical dynamics implies abrupt changes in the behavior of
the system avaliable phase space volume, which can be considered as 
the semiclassical approximation of the quantum density of states \cite{Gutzwiller2013}.
For the anharmonic LMG model, the avaliable phase space volume is given by \cite{Gamito2022}
\be\label{DoS}
 \rho_{sc}(\varepsilon)=\frac{1}{4\pi}\int dpdq\delta[\varepsilon-\mathcal{H}_c(p,q)],
\ee
which can be analytically calculated via the properties of the delta function, 
as performed in Ref.~\cite{Gamito2022}. 

The heat map plot in Fig.~\ref{EgDoS}(a) shows $\rho_{sc}(\varepsilon)$ as a function of $\gamma$
and $\varepsilon$ for $\alpha=0.5$.
One can clearly appreciate how the behavior of $\rho_{sc}(\varepsilon)$ exhibts maximums along
the critical energies $\varepsilon_{c,1}$ and $\varepsilon_{c,2}$.
The singularities in the density of states are more visible in the 
dependence of $\rho_{sc}(\varepsilon)$ on $\gamma$ and $\varepsilon$, respectively.
Figures \ref{EgDoS}(b) and \ref{EgDoS}(c) are the respective plots 
for the variation of $\rho_{sc}$ with $\gamma$ and $\varepsilon$ along 
the lines $\varepsilon=0.2$ and $\gamma=0.7$.
We clearly observe that $\rho_{sc}(\varepsilon)$ shows a sharp peak at the critical energies of ESQPTs.
In fact, for the anharmonic LMG model with the classical counterpart has one degrees of freedom,
it was demonstrated that around the ESQPT critical energy $\rho_{sc}(\varepsilon)$ bears the logarithmic divergence, so that $\rho_{sc}(\varepsilon)\propto-\ln|\varepsilon-\varepsilon_{c}|$
\cite{Stransky2014,Cejnar2021}. 
Moreover, we also see an excellent agreement between the numerical results of 
$\tilde{\rho}(\varepsilon)$ and $\rho_{sc}(\varepsilon)$.
This confirms that both ESQPTs are captured by the logarithmic divergences in the density of states
as the system size goes to infinite. 

In the following section, we investigate the implications of these ESQPTs in the system
nonequilibrium thermodynamics and discuss how to reveal 
them by means of the entropy of the quantum work distribution.

 \begin{figure}
  \includegraphics[width=\columnwidth]{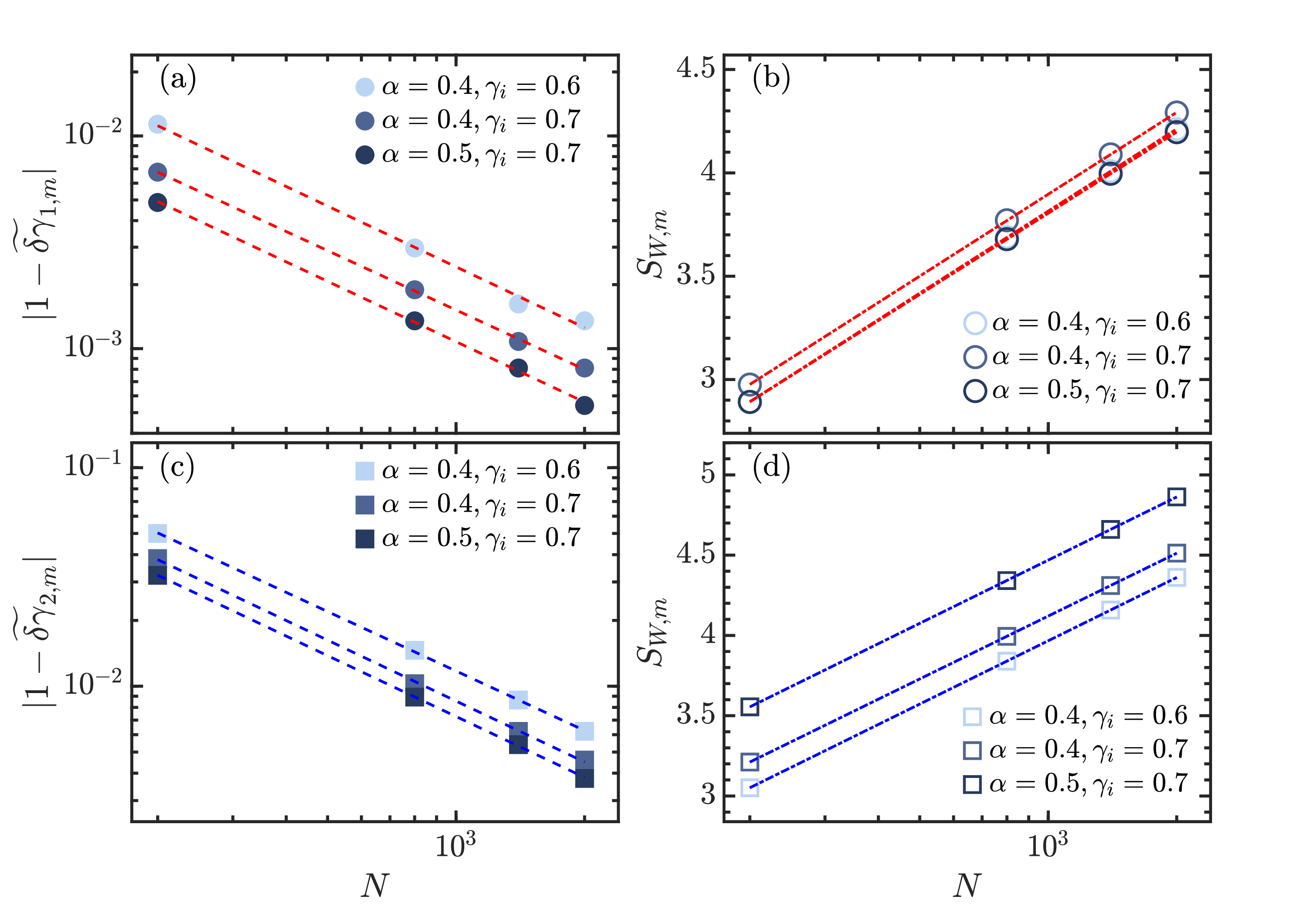}
  \caption{(a) Distances of the rescaled quenching strength $\widetilde{\delta\gamma}_{1,m}$ 
  where the maximal $S_W$ occurs to its critical value $1$ versus system size $N$ 
  with several $\alpha$ and $\gamma_i$ for the first ESQPT.
  (b) Maximum values of $S_W$, $S_{W,m}$,  as a function of $N$ for different values of $\alpha$ and
  $\gamma_i$ of the first ESQPT.
  (c) Distances between $\widetilde{\delta\gamma}_{2,m}$, where $S_W$ reaches its maximal value, 
  and $1$ as a function of $N$ for several $\alpha$ and $\gamma_i$ of the second ESQPT. 
 (d) Maximal values of $S_W$, $S_{W,m}$, versus $N$ for different $\alpha$ and $\gamma_i$ 
 of the second ESQPT.
 The dashed lines in panels (a) and (c) corresponde to the power law scaling $N^{-\mu}$,
 while the dotted lines in panels (b) and (d) represent $\nu\ln N$.
  The explicit values of $\mu$ and $\nu$ are given in Table \ref{Expvalues}.
  All quantities are unitless.}
  \label{ScEntropy}
 \end{figure}

\begin{table*}
\caption{\label{Expvalues}Scaling exponents $\mu$ and $\nu$ for the cases plotted
in Fig.~\ref{ScEntropy}.}
\begin{ruledtabular}
\begin{tabular}{cccccc}
 \multicolumn{3}{c}{1st ESQPT}&\multicolumn{3}{c}{2nd ESQPT}\\
  \cline{1-3}  \cline{4-6}\\
 $\alpha=0.4, \gamma_i=0.6$&$\alpha=0.4, \gamma_i=0.7$&$\alpha=0.5, \gamma_i=0.7$
& $\alpha=0.4, \gamma_i=0.6$ &$\alpha=0.4, \gamma_i=0.7$  & $\alpha=0.5, \gamma_i=0.7$\\
\cline{1-3}  \cline{4-6}\\
\makecell{$\mu=0.9496$\\ $\nu=0.5732$}&\makecell{$\mu=0.9280$\\ $\nu=0.5719$}
&\makecell{$\mu=0.9432$\\ $\nu=0.5670$}
&\makecell{$\mu=0.9043$\\$\nu=0.5688$}&\makecell{$\mu=0.9262$\\ $\nu=0.5651$} &
\makecell{$\mu=0.9251$\\ $\nu=0.5679$} \\
\end{tabular}
\end{ruledtabular}
\end{table*}

\section{Results}\label{FourthS}

To analyse how two ESQPTs in the anharmonic LMG model affect the nonequilibrium thermodynamics
and their associated signatures through the work distribution entropy, 
we keep the value of $\alpha$ fixed and consider a sudden quench protocol.
The initial state of the system is prepared in an eigenstate $|\psi_n^i\ra$ 
of $H_i=H(\gamma_i)$ with energy $E_n^i$. 
At $t=0$, a sudden quench takes place which changes $\gamma$ 
from $\gamma_i$ to $\gamma_f=\gamma_i+\delta\gamma$.
The final Hamiltonian of the system is $H_f=H(\gamma_f)=\sum_kE_k^f|\psi_k^f\ra\la\psi_k^f|$
where $|\psi_k^f\ra$ is the $k$th eigenstate of $H_f$ with eigenvalue $E_k^f$.
As $\rho_i=|\psi_n^i\ra\la\psi_n^i|$ is the $n$th eigenstate of $H_i$ and the unitary opertor
$U=\mathbbm{1}$ for the sudden quenche process, 
the work distribution in Eq.~(\ref{DefPw}) reduces to
\be\label{PDW}
  P(W)=\sum_kp_{k,n}\delta[W-(E_k^f-E_n^i)],
\ee
where the joint probability in Eq.~(\ref{JointPrb}) simplifies to 
$p_{k,n}=|\la\psi_k^f|\psi_n^i\ra|^2$, namely, 
the transition probabilities between the inital and final states.
Then, the entropy defined in Eq.~(\ref{EntropyPw}) is given by
\be \label{Pdwentropy}
  S_W=-\sum_W P(W)\ln[P(W)]=-\sum_k p_{k,n}\ln p_{k,n}.
\ee
Obviouly, the entropy $S_W$ now
varies in the interval $S_W\in[ 0,\ln\mathcal{D}]$ with $S_W=0$ corresponds to the 
deterministic work and $S_W=\ln\mathcal{D}$ implies $P(W)$ is uniform.
Here, $\mathcal{D}$ is the Hilbert space dimension of the system.

Since we aim to reveal the effects and characterizations of the ESQPTs, 
it is necessitated to take the system passes through the critical energies of ESQPTs.
This is achevied by tuning the quenching strength $\delta\gamma$, owing to the dependence of 
the energy in the post-quenched system on $\delta\gamma$ value. 
The critical quenching is defined as the one that takes the post-quenched system
to the ESQPT criticl energy, and denoted by $\delta\gamma_c$.

The critical quenching of an ESQPT can be obtained by using 
the mean field (semiclassical) approach.
For the first ESQPT with the initial state is given by the ground state $|\psi_0^i\ra$ of $H_i$,
the critical quenching, $\delta\gamma_{c,1}$, can be written as
\be\label{CrtFirst}
  \delta\gamma_{c,1}=-\frac{(3\gamma_i-1)(2\gamma_i-\alpha)}{2(3\gamma_i-3\alpha+1)},
\ee
with $1/3\leq\gamma_i\leq1$.
On the other hand, it was demonstrated that it is impossible to approach the second ESQPT 
when the system is initially in the ground state of $H_i$ \cite{Khalouf2022a}.
Alternatively, to reach the critical energy of the second ESQPT, the initial state should be 
set as the highest excited eigenstate, denoted by $|\psi_{n^\ast}^i\ra$, of $H_i$ \cite{Khalouf2022a}. 
Then, one can find that the critical quenching, $\delta\gamma_{c,2}$, 
of the second ESQPT is given by
\be\label{CrtSecond}
  \delta\gamma_{c,2}=\frac{4\alpha(1-\gamma_i-\alpha)-(1-\gamma_i)^2}{2(2\alpha+\gamma_i-1)},
\ee
where, again, $1/3\leq\gamma_i\leq1$.
We would like to emphasize that the conclusions in the present work are independent of the value of
$\gamma_i$ as long as $\gamma_i\in[1/3, 1]$.

In Fig.~\ref{PdfWork}, we plot $P(W)$ for different rescaled quenching strengths, 
$\widetilde{\delta\gamma}_a=\delta\gamma/\delta\gamma_{c,a}$ with $a=1,2$ for the first
[Figs.~\ref{PdfWork}(a)-\ref{PdfWork}(c)] and second [Figs.~\ref{PdfWork}(d)-\ref{PdfWork}(f)]
ESQPTs, with $\gamma_i=0.7$, $\alpha=0.5$ and $N=2j=800$.
Overall, the behavior of $P(W)$ clearly unviels the two ESQPTs 
at $\delta\gamma_{c,1}$ and $\delta\gamma_{c,2}$.
For both ESQPTs, when $\widetilde{\delta\gamma}_a<1$ 
the work distribution $P(W)$ has small support and shows 
significant population around the work value given by the energy difference 
between the ground states of $H_f$ and $H_i$, 
as illustrated in Figs.~\ref{PdfWork}(a) and \ref{PdfWork}(d).   
Conversely, as evidenced in Figs.~\ref{PdfWork}(c) and \ref{PdfWork}(f), 
the support of $P(W)$ undergoes a remarkable increase for
quenches that above the critical ones, i.~e. $\widetilde{\delta\gamma}_a>1$.
The particular dip observed in $P(W)$ for the critical quenches $\widetilde{\delta\gamma}_a=1$ 
[see Figs.~\ref{PdfWork}(b) and \ref{PdfWork}(e)] not only marks the presence of ESQPTs, 
but also reflects the complexity of $P(W)$ at the ESQPTs critical points. 
However, we see that the work distribution is very regular when the quenching strength
far away from the critical value for both ESQPTs.

Above observed features of $P(W)$ imply that the entropy of $P(W)$ should exhibit 
a drastic change as the system passes through the critical points of the two ESQPTs.
To see this, we plot $S_W$ as a function of quenching strength and $\gamma_i$ for two ESQPTs in
Figs.~\ref{PWentropy}(a) and \ref{PWentropy}(b).
For both ESQPTs, one can clearly see that $S_W$ exhibits obvious different behaviors in 
different phases of an ESQPT. 
In particular, the entropy $S_W$ is maximum around the critical point in both transitions.  
These properties of $S_W$ are more visible in Figs.~\ref{PWentropy}(c) and \ref{PWentropy}(d), 
where we show the dependence of $S_W$ on the quenching strength with fixed $\gamma_i$ for
two ESQPTs.

The peak displayed in the behavior of $S_W$ implies that it succinctly reveals the ESQPTs in the
anharmonic LMG model and acts as a finite size precursor of an ESQPT.
Hence, one can expect that in both transitions the location of the maximal entropy tends to the
critical point and the entropy diverges in the thermodynamic limit $N\to\infty$.
This is confirmed by Fig.~\ref{ScEntropy}, where we demonstrate how the 
position of the maximal $S_W$ with respect to the critical value as well as 
the maximum value of  $S_W$ evolve with the system size $N$ 
for different $\alpha$ and $\gamma_i$ cases in both transitions.
Moreover, we find that the decrease of the distances between the location 
of the maximal $S_W$ and the critical point with the increase of $N$ follows power law
\be
  |1-\widetilde{\delta\gamma}_{a,m}|\propto N^{-\mu},
\ee
where $a=1,2$ and $\widetilde{\delta\gamma}_{a,m}=\delta\gamma_{a,m}/\delta\gamma_{c,a}$
with $\delta\gamma_{a,m}$ denotes the position of the maximal $S_W$ for the first ($a=1$) and/or
second ($a=2$) ESQPT.
Additionally, the best fit of the data shows that for both ESQPTs the divergence of 
the maximum value of $S_W$, denoted by $S_{W,m}$, is well captured by
\be
   S_{W,m}\propto\nu\ln N.
\ee
The values of the scaling exponents $\mu$ and $\nu$ for the two ESQPTs are 
shown in Table \ref{Expvalues}.
We see that for both transitions the exponent $\nu$ is almost independent of the values 
of $\alpha$ and $\gamma_i$ and approximately given by $\nu\approx0.57$, 
while the value of $\mu$ is different for two transitions and 
varies with $\alpha$ and $\gamma_i$ as well.

So far, we have focused on to investigate the impacts and characterizations of the ESQPTs 
by means of the entropy of the quantum work distribution in the anharmonic LMG parameter space.
However, the prominent signature of ESQPTs is the singularity 
in the density of states at the critical energy. 
Therefor, it is also necessary to explore how the entropy of the quantum work distribution varies
with different energy levels. 

To this end, we still consider the sudden quench process.
However, as we are currently interested in the
energy denpendence of the entropy of the work distribution, we fixed $\delta\gamma=0.001$ and
study the properties of the work distribution and its entropy for different eigenstates 
with fixed $\gamma_i$.

\begin{figure}
  \includegraphics[width=\columnwidth]{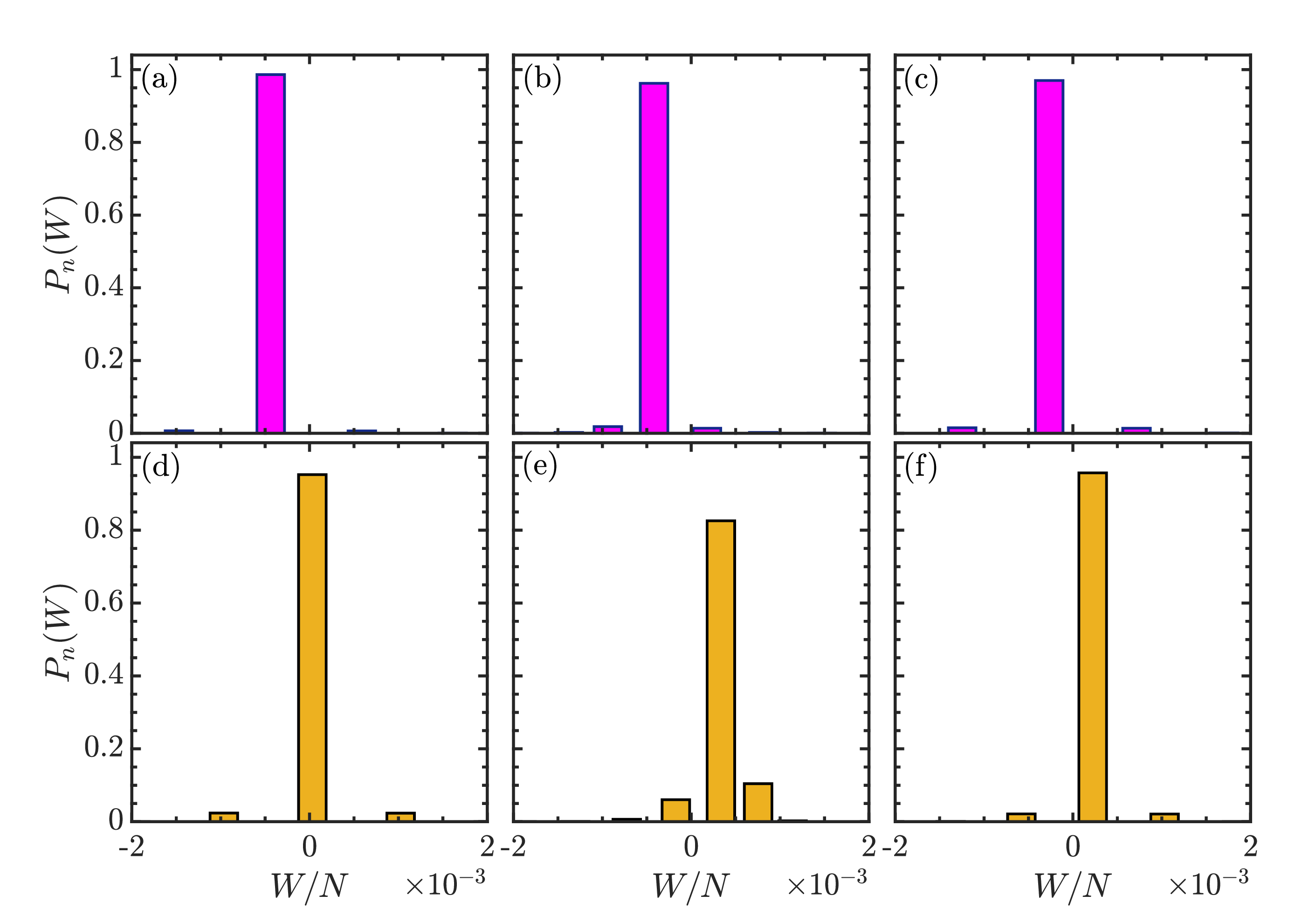}
  \caption{Work distribution $P_n(W)$ of the $n$th eigenstate $|\psi_n\ra$ for
  (a) $n=79, \varepsilon_n\approx0.094$, (b) $n=129, \varepsilon_n\approx\varepsilon_{c,2}=0.1361$,
  (c) $n=179, \varepsilon_n\approx0.179$, (d) $n=279, \varepsilon\approx0.2818$, 
  (e) $n=344, \varepsilon_n\approx\varepsilon_{c,1}=0.3361$, 
  and (f) $n=371, \varepsilon_n\approx0.3544$.
  Here, $\varepsilon_n=(E_n-E_0)/N$ with $E_0$ is the ground state energy, $\varepsilon_{c,1}$ and 
  $\varepsilon_{c,2}$ are, respectively, given by Eqs.~(\ref{CrtEg1}) and (\ref{CrtEg2}).
 Other parameters are: $\gamma_i=0.7$, $\alpha=0.5$, and $N=2j=800$.
  All quantities are unitless.}
  \label{PdfWEg}
 \end{figure}

In Fig.~\ref{PdfWEg}, we report the work distribution of $n$th eigenstate, denoted by $P_n(W)$, 
for several excitation eigenlevels with $\gamma_i=0.7$, $\alpha=0.5$, and the system size $N=2j=800$,
for the two ESQPTs.
Due to the very small value of $\delta\gamma$, the amount of work that is injected or extracted during
the quench process is also very small.
Nevertheless, for both transitions, obvious differences in the behavior of $P_n(W)$ 
between two phases of an ESQPT can be observed.
At the critical energies of two ESQPTs, the highest density of states means that 
the critical eigenstates are very sensitive to the small perturbations.
As a consequence, the work distribition $P_n(W)$ shows a lower peak compared to 
the cases that far away from the critical energy, 
as seen in Figs.~\ref{PdfWEg}(b) and \ref{PdfWEg}(e).

\begin{figure}
  \includegraphics[width=\columnwidth]{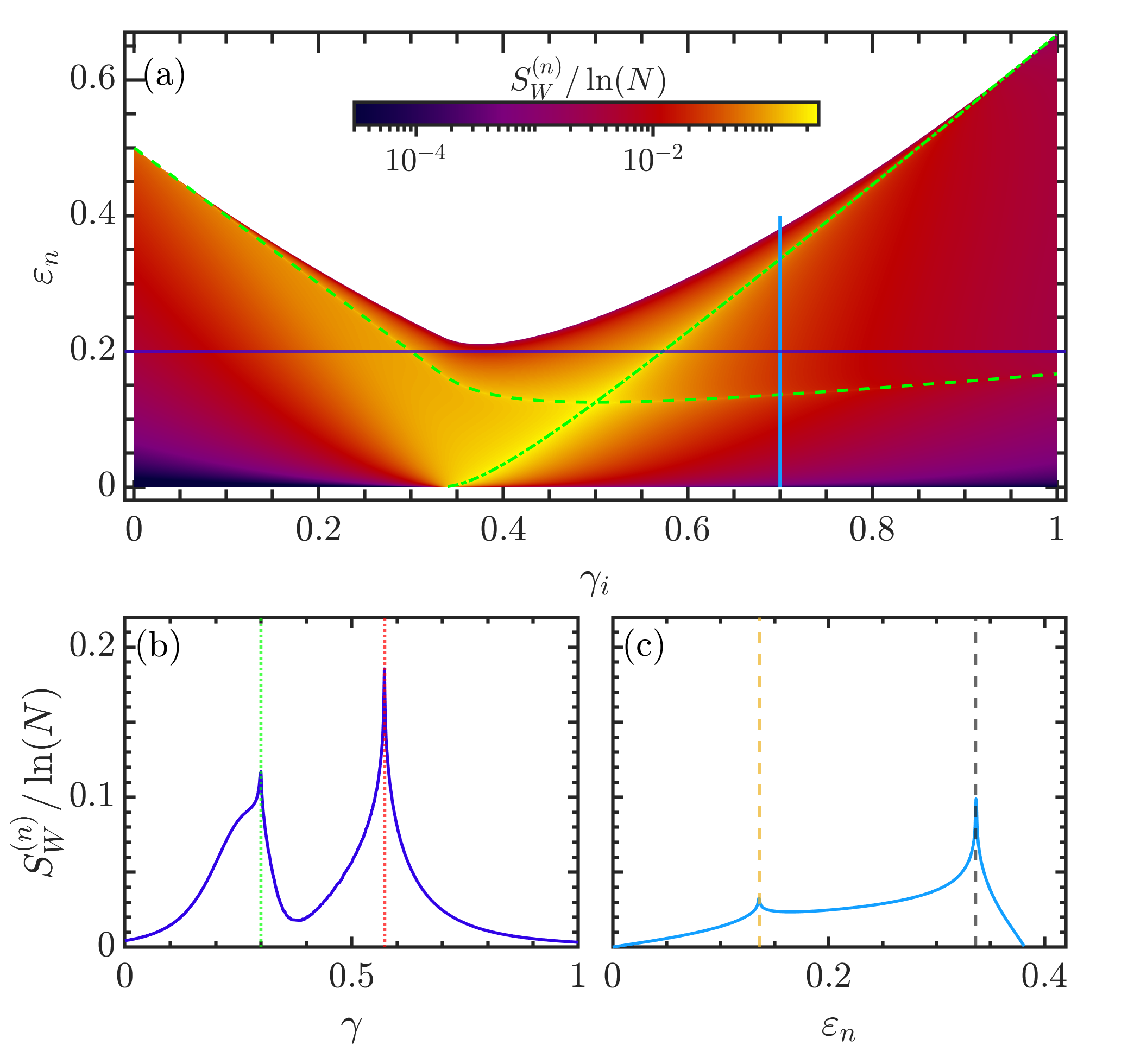}
  \caption{(a) Rescaled entropy of $P_n(W)$, $S_W^{(n)}/\ln N$, as a function of $\gamma_i$ and
  $\varepsilon_n=(E_n-E_0)/N$. Here, $E_0$ is the ground state energy.
  The green dot dashed and dashed curves mark the critical energies 
  $\varepsilon_{c,1}$ and $\varepsilon_{c,2}$, 
  given by Eqs.~(\ref{CrtEg1}) and (\ref{CrtEg2}), respectively.
  The horizatonal and vertical solid lines denote $\varepsilon_n=0.2$ and $\gamma_i=0.7$.
  (b) $S_W^{(n)}/\ln N$ as a function of $\gamma_i$ along $\varepsilon_n=0.2$.
  The vertical red and green dotted lines mark the critical values of $\gamma_i$ obtained from
  $\varepsilon_{c,1}=\varepsilon_{c,2}=0.2$.
  (c) $S_W^{(n)}/\ln N$ versus $\varepsilon_n$ with $\gamma_i=0.7$.
  The vertical gray and orange dashed lines correspond to $\varepsilon_{c,1}$ and
  $\varepsilon_{c,2}$.
  Other parameters are: $\alpha=0.5$, $\delta\gamma=0.001$, and $N=2j=800$.
  All quantities are unitless.}
  \label{EntropyEg}
 \end{figure}

The behaviors of $P_n(W)$ demonstrated in Fig.~\ref{PdfWEg} indicate that
the entropy of $P_n(W)$, denoted by $S_W^{(n)}$, would be maximized 
at the ESQPT critical energy.
We plot in Fig.~\ref{EntropyEg}(a) how the entropy $S_W^{(n)}$ evolves
with $\gamma_i$ and excitation energies $\varepsilon_n$.
We note that the overall behavior of $S_W^{(n)}$ is very similar to $\rho_{sc}(\varepsilon)$,
as seen by comparing Fig.~\ref{EntropyEg}(a) with Fig.~\ref{EgDoS}(a).
Importantly, one can clearly observe that the entropy $S_W^{(n)}$ has a maximum along
the critical energies of two ESQPTs.
This means that the underlying  ESQPTs in the anharmonic LMG model have strong
impacts on its nonequilibrium thermodynamical properties and the entropy 
behaves as a witness of both ESQPTs.
To further verify this statement, we show the dependence of $S_W^{(n)}$ 
on $\gamma_i$ with fixed $\varepsilon_n$ and on $\varepsilon_n$ with fixed $\gamma_i$ in 
Figs.~\ref{EntropyEg}(b) and \ref{EntropyEg}(c), respectively.
Two visible sharp peaks near the critical points of two ESQPTs in the behavior of $S_W^{(n)}$
allow us to conclude that the presence of ESQPTs in the 
anharmonic LMG model can be reliably probed 
by the entropy of the quantum work distrbution.
Hence, the usefuless of the entropy of the quantum work distribution is twofold.
On the one hand, it allows us to characterize 
the ESQPT affects on the nonequilibrium thermodynamical 
properties in a quantum system. 
On the other hand, and consequently it can be utilized to reveal the presence of ESQPTs.

\section{Conclusion} \label{FivthS}

In conclusion, using the entropy of the quantum work distribution,
we have investigated how the the nonequilibrium thermodynamics is affected by
the ESQPT as well as associated critical signatures in the anharmonic LMG model.
As a generalization of the well know LMG model, the anharmonic LMG model includes
an anharmonic term in the LMG Hamiltonian.
As a consequence, in addition to the known ESQPT which associated with the ground state QPT
and has been observed in the LMG model, a new ESQPT that is 
induced by the anharmonic term is present in the anharmonic LMG model. 

To understand the new ESQPT, we have studied the classical limit of the model.
We shown that although the physical origines of the two ESQPTs are different, both of them are 
signified by the logarithmic divergence of the density of states at their critical energies.
We have performed the stability analysis of the classical Hamiltonian and deduced 
the explicit form of the critical energies of the two ESQPTs.

The entropy of the quantum work distribution measures the complexity of the distribution.
By focusing on the sudden quenche process,
we have demonstrated that the entropy is acutely sensitive to the two ESQPTs, resulting in
a significant change of the property of the entropy 
when we straddled the critical points of the two ESQPTs.
In particular, we have shown that the presence of the two ESQPTs can be clearly revealed by the peaks
in the behavior of the entropy.
Further scaling analyses of the maximal entropy shows that it can be recongnized as a finte size 
precursor for both ESQPTs.
Moreover, we have demonstrated that the entropy is also capable of diagnosing the emergence of ESQPTs
in the energy space and can be used to detect the critical energies of the two ESQPTs.

Our study extends the investigation of the implications of the two ESQPTs in the 
anharmonic LMG model on system static and dynamical 
properties \cite{Gamito2022,Khalouf2022a} to the noneqilibrium thermodynamics 
and provides further verification of the usefuless of the quantum work
distribution entropy in understanding and detecting different phase transtions
in quantum many body systems. 
We would like to point out that our main conclusions, despite of
building on the anharmonic LMG model,
are general for the ESQPTs that are characterized by the 
logarithmic divergence of the density of states. 
Whether the results in this work are still hold for other kinds of ESQPTs, 
such as the one defined as the nonanlytical in the first drivative of the density of states,
remains an open question and deserves further exploration. 
Another possible extension of the present work would be 
to elucidate whether the finite size scaling analyses of the entropy 
can help us to classify various ESQPTs.

Finally, the quantum work distribution based on the two point measurement scheme 
has been experimentally measured in several platforms
\cite{Dorner2013,Mazzola2013,Batalhao2014,An2015}.
We therefore hope that our findings could open up a promising way for the use of 
the quantum work distribution entropy in the experimental studies of ESQPTs.

 \acknowledgements
 
 Q.~W. acknowledges support from
 the Slovenian Research and Innovation Agency (ARIS) 
 under Grant Nos. J1-4387 and P1-0306; 
 This work was supported by 
 the Zhejiang Provincial Nature Science Foundation 
 under Grant Nos. LQ22A040006 and LY20A050001;
 the National Science Foundation of China under Grant No. 11805165.

\bibliographystyle{apsrev4-1}
\bibliography{WorkALMG}

\end{document}